\documentclass[10pt,twocolumn,letterpaper]{article}

\usepackage{cvpr} 
\usepackage[accsupp]{axessibility} 

\usepackage{booktabs} 
\usepackage{tabularx}
\usepackage{subcaption}
\usepackage{multirow}
\usepackage{amsmath}
\usepackage{mathrsfs}
\usepackage{graphicx}
\usepackage{threeparttable}

\usepackage{algorithm}
\usepackage[noend]{algpseudocode}

\usepackage{bbding}
\usepackage{tabularx}

\usepackage[dvipsnames]{xcolor}

\definecolor{cvprblue}{rgb}{0.21,0.49,0.74}
\usepackage[pagebackref,breaklinks,colorlinks,citecolor=cvprblue]{hyperref}

\def\confName{CVPR}
\def\confYear{2024}

\title{DiffSHEG: A Diffusion-Based Approach for Real-Time Speech-driven Holistic 3D Expression and Gesture Generation}

\author{
Junming Chen$^{1,2}$\thanks{The work was done during Junming's internship at the International Digital Economy Academy.} \quad Yunfei Liu$^1$  \quad Jianan Wang$^1$ \quad Ailing Zeng$^1$ \quad Yu Li$^1$\thanks{Corresponding authors.}  \quad Qifeng Chen$^2$\footnotemark[2]  \\
  $^1$International Digital Economy Academy \quad $^2$Hong Kong University of Science and Technology\\
{\tt \small \{jchenfo, cqf\}@ust.hk \quad \{liuyunfei, wangjianan, zengailing, liyu\}@idea.edu.cn} \\
\texttt{\href{https://jeremycjm.github.io/proj/DiffSHEG}{https://jeremycjm.github.io/proj/DiffSHEG}}
}

\newcommand{\groundtruth}[1]{\text{#1}}

\begin{document}
\maketitle



\begin{abstract}
We propose \textbf{DiffSHEG}, a \textbf{Diff}usion-based approach for \textbf{S}peech-driven \textbf{H}olistic 3D \textbf{E}xpression and \textbf{G}esture generation with arbitrary length. While previous works focused on co-speech gesture or expression generation individually, the joint generation of synchronized expressions and gestures remains barely explored. To address this, our diffusion-based co-speech motion generation transformer enables uni-directional information flow from expression to gesture, facilitating improved matching of joint expression-gesture distributions. Furthermore, we introduce an outpainting-based sampling strategy for arbitrary long sequence generation in diffusion models, offering flexibility and computational efficiency. Our method provides a practical solution that produces high-quality synchronized expression and gesture generation driven by speech. Evaluated on two public datasets, our approach achieves state-of-the-art performance both quantitatively and qualitatively. Additionally, a user study confirms the superiority of DiffSHEG over prior approaches. By enabling the real-time generation of expressive and synchronized motions, DiffSHEG showcases its potential for various applications in the development of digital humans and embodied agents.
\vspace{-0.45cm}
\end{abstract}

\begin{figure}[t]
\begin{center}
\includegraphics[width=\linewidth]{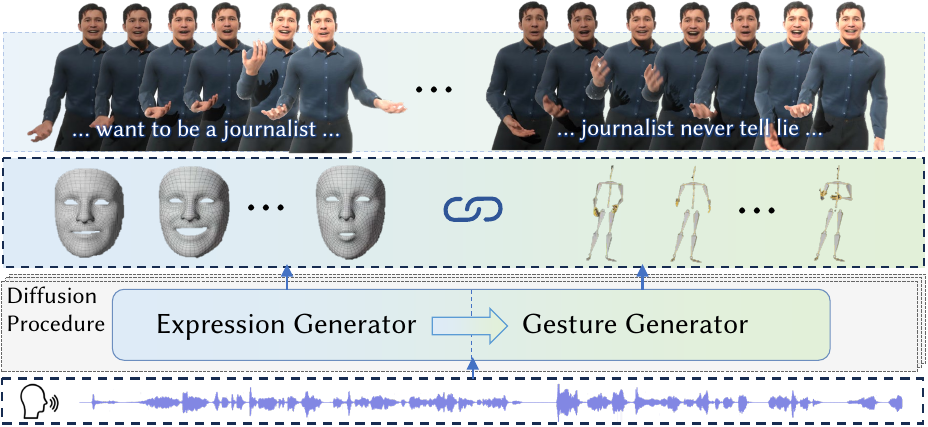}
\end{center}
\vspace{-0.45cm}
   \caption{\textbf{DiffSHEG} is a unified co-speech expression and gesture generation system based on diffusion models. It captures the joint expression-gesture distribution by enabling the uni-directional information flow from expression to gesture inside the model.}
\label{fig:teaser}
\vspace{-0.45cm}
\end{figure}

\section{Introduction}
\label{sec:intro}
Non-verbal cues such as facial expressions, body language, and hand gestures play a vital role in effective communication alongside verbal language~\cite{van1998persona, goldin1999role}. Speech-driven gesture and expression generation has gained significant interest in applications like the metaverse, digital human development, gaming, and human-computer interaction. Generating synchronized and realistic gestures and expressions based on speech is key to bringing virtual agents and digital humans to life in virtual environments. 

Existing research on co-speech motion synthesis has focused on generating either expressions or gestures independently. Rule-based approaches~\cite{Kipp2004_Gesture, huang2012robot, softbank2018naoqi, cassell2004beat, wagner2014gesture} were prevalent initially, but recent advancements have leveraged data-driven techniques using deep neural networks. However, co-speech gesture generation poses a challenge due to its inherently many-to-many mapping. State-of-the-art methods have explored generative models such as normalizing flow models~\cite{ye2022audio}, VQ-VAE~\cite{ao2022rhythmic}, GAN~\cite{habibie2021learning} and Diffusion models~\cite{zhu2023taming, Ao2023GestureDiffuCLIP, alexanderson2023listen, ijcai2023p650}. 
These approaches have made progress in improving synchronization and diversifying generated gestures. However, none of them specifically target the co-speech generation of both expressions and gestures simultaneously.

Recently, some works have aimed to generate co-speech holistic 3D expressions and gestures~\cite{talkshow, habibie2021learning}. These methods either combine independent co-speech expression and gesture models~\cite{talkshow} or formulate the problem as a multi-task learning one~\cite{habibie2021learning}. 
However, these approaches separate the generation process of expressions and gestures, neglecting the potential relationship between them. This can lead to disharmony and deviation in the joint expression-gesture distribution. Additionally, deterministic CNN-based models~\cite{habibie2021learning} may not be well-suited for approximating the many-to-many mapping inherent in co-speech gesture generation.

\begin{figure*}[t]
\begin{center}
\includegraphics[width=\linewidth]{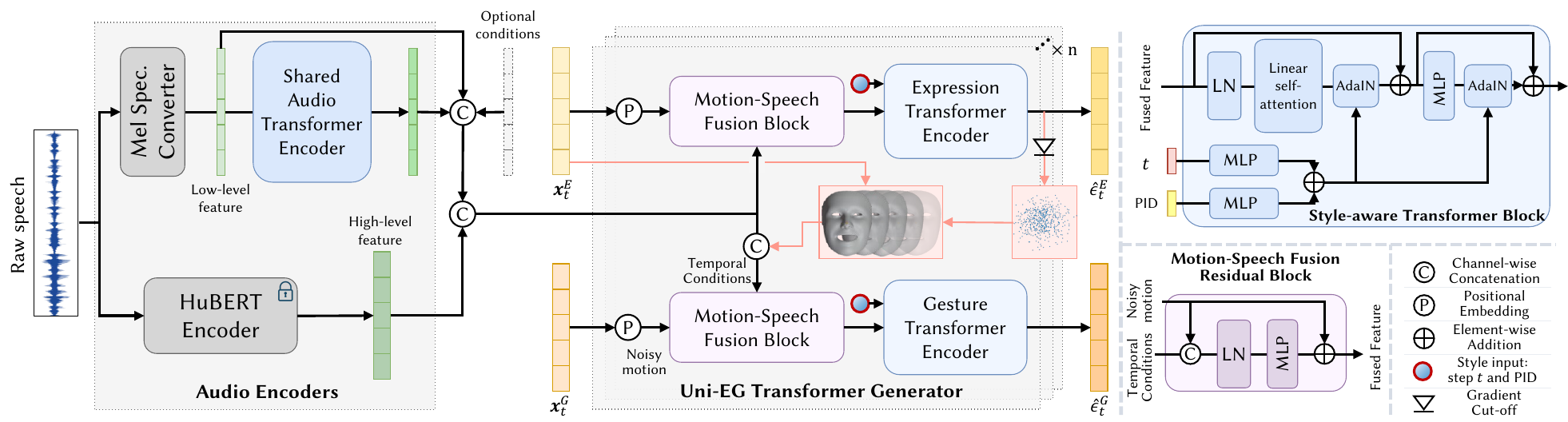}
\end{center}
\vspace{-0.5cm}
   \caption{\textbf{DiffSHEG framework overview}. \textbf{Left}: Audio Encoders and UniEG-Transformer Generator. Given an audio clip, we encode the audio into a low-level feature Mel-Spectrogram and a high-level HuBERT feature. An audio encoder learns a mid-level representation of speech. The audio features are concatenated with other optional temporal conditions and then fed into the UniEG Transformer Denoiser. The denoising block fuses the conditions with noisy motion at diffusion step t and feeds it into style-aware transformers to get the predicted noises. The uni-directional condition flow is enforced from expression to gesture for joint distribution learning. \textbf{Right}: The detailed architecture of style-aware Transformer encoder and motion-condition fusion residual block.}
\label{fig:framework}
\vspace{-0.45cm}
\end{figure*}

In this work, we propose \textbf{DiffSHEG}, a unified \textbf{D}iffusion-based \textbf{H}olistic \textbf{S}peech-driven \textbf{E}xpression and \textbf{G}esture generation framework, illustrated in Figure~\ref{fig:teaser}. 
To capture the joint distribution, DiffSHEG utilizes diffusion models~\cite{ddpm} with a unified expression-gesture denoising network. 
As shown in Figure~\ref{fig:framework}, our denoising network consists of two modules, an audio encoder, and a Transformer-based Uni-direction Expression-Gesture (\textbf{UniEG}) generator, which has a unidirectional flow from expression to gesture. The proposed framework ensures a natural temporal alignment between speech and motion, leveraging the relationship between expressions and gestures. Furthermore, we introduce a Fast Out-Painting-based Partial Autoregressive Sampling (\textbf{FOPPAS}) method to synthesize arbitrary long sequences efficiently. FOPPAS enables real-time streaming sequence generation without conditioning on previous frames during training, providing more flexibility and efficiency. 

Our contributions can be summarized as follows:
\textbf{(1)} We develop a unified diffusion-based approach for speech-driven holistic 3D expression and gesture generation framework: DiffSHEG. It is the first attempt to explicitly model the joint distribution of expression and gesture. 
\textbf{(2)} To better capture the expression-gesture joint distribution, we design a uni-directional expression-gesture (UniEG) Transformer generator, which enforces the uni-directional condition flow from expression to gesture generator. 
\textbf{(3)} We introduce FOPPAS, a fast out-painting-based partial autoregressive sampling method. FOPPAS enables real-time generation of arbitrary long smooth motion sequences using diffusion models. It achieves over 30 FPS on a single Nvidia 3090 GPU and can work with streaming audio.
\textbf{(4)} We evaluate our method on two new public datasets and achieve state-of-the-art performance quantitatively and qualitatively. User studies validate the superiority of our method in terms of motion realism, synchronism, and diversity.

\section{Related Work}

\noindent \textbf{Co-speech Expression Generation.}
Co-speech Expression Generation, also known as talking head/face generation, has been a topic of interest in the field. Early related methods primarily relied on rule-based procedural approaches~\cite{edwards2016jali, massaro2012, taylor2012dynamic, xu2013practical}, which required significant manual effort despite offering explicit and flexible control.
With the advancements in deep learning, data-driven methods~\cite{cao2005expressive,liu2015video,karras2017audio,taylor2017deep,pham2018end,cudeiro2019capture,hussen2020modality,zhang2022sadtalker,shen2023difftalk,Liu_2023_ICCV} have emerged, focusing on generating images that correspond to audio input. However, these methods often produce distorted pixels and inconsistent 3D results.
On another front, some approaches~\cite{karras2017audio,richard2021meshtalk} propose animating 3D faces. More recently, FaceFormer~\cite{fan2022faceformer} and CodeTalker~\cite{xing2023codetalker} have utilized Transformer models to animate 3D facial vertices using designed attention maps for temporal alignment. While lip movement is closely tied to speech, most data-driven methods for expression animation design deterministic mappings, lacking diversity in eye movement and facial expressions.
In contrast to these methods, our approach aims to leverage generative models to generate diverse facial expressions with precise lip movements driven by speech.

\noindent \textbf{Co-speech Gesture Generation.}
The speech-driven gesture generation follows a similar path as face animation: from rule-based~\cite{Kipp2004_Gesture, huang2012robot, softbank2018naoqi, cassell2004beat, wagner2014gesture} to data-driven methods~\cite{kopp2006towards, kucherenko2020gesticulator, bhattacharya2021speech2affectivegestures, ao2022rhythmic}. The early data-driven methods utilize statistical models to learn speech-gesture mapping. 
After that, deterministic models such as multilayer perceptron (MLP)~\cite{kucherenko2020gesticulator}, convolutional networks~\cite{habibie2021learning}, recurrent neural networks~\cite{hasegawa2018evaluation, yoon2019robots, yoon2020speech, bhattacharya2021speech2affectivegestures, liu2022learning}, and Transformers~\cite{9417647} are explored.
Since speech-to-gesture mapping involves a many-to-many mapping, numerous state-of-the-art methods utilize generative models like normalizing flow~\cite{ye2022audio}, GAN~\cite{habibie2021learning} and diffusion-based models~\cite{alexanderson2023listen, zhu2023taming, ijcai2023p650}. 
Recently, DiffGesture~\cite{zhu2023taming} and DiffuseStyleGesture~\cite{ijcai2023p650}  first applied diffusion in gesture generation. However, both frameworks require initial gestures during training for long sequence generation, which lacks flexibility and diversity compared to our FOPPAS. LDA~\cite{alexanderson2023listen} utilizes Conformer as the diffusion base network and takes advantage of translation-invariant positional embedding to sample longer sequences in a go. Nevertheless, it cannot work with streaming audio and may not generalize well on very long sequences. 
%



\noindent \textbf{Holistic Co-Speech Expression and Gesture Generation.}
There are some recent methods~\cite{habibie2021learning, talkshow} exploring the holistic joint generation of expression and gesture. 
Habibie et al.~\cite{habibie2021learning} first try to synthesize 3D facial and gesture motion at the same time. They propose a CNN-based framework that shares the same speech encoder and uses three decoders to generate distinct whole-body keypoints (i.e., for face, hand, body), followed by a discriminator to provide adversarial loss during training. However, this method learns deterministic mapping, which violates the nature of many-to-many mapping from speech to gesture motion. 
Yi et al.~\cite{talkshow} use an encoder-decoder network with pre-trained Wav2Vec~\cite{baevski2020wav2vec} audio feature for expression generation and VQ-VAE for gesture generation. Nevertheless, they generate the expressions and gestures separately, overlooking the joint distribution between them. Moreover, methods based on VQ-VAE are constrained by a finite set of motion tokens, potentially limiting their ability to generate diverse and agile motions.

\section{Method}
\label{sec:method}

How to model the expression-gesture joint distribution can be key to this task.
A na\"ive way is directly concatenating the two vectors together and passing them to deep networks simultaneously, such that the features of both can be shared by each other. However, we empirically find this does not work well and leads to sub-optimal results in our experiment. 
We hypothesize that the information flow from gesture to expression would interfere with the mapping from speech to expression. Intuitively, expressions can serve as cues for gestures since the gestures usually behave consistently with the emotions and speech inferred from expression~\cite{emotion_recognition} and lip movement~\cite{vid2speech}. 
On the contrary, gestures can hardly affect expressions, especially the lips, which have a strong and exclusive correlation with speech. 
Imagine when you express surprise, you may have widened eyes, opened mouth, and raised eyebrows, with frozen or startled gestures such as clasping hands and leaning backward.
Therefore, we propose a unified framework based on diffusion models with \textbf{uni-directional information flow from expression to gesture} for unified co-speech expression and gesture generation to capture their joint distributions. At inference time, to generate arbitrary-long motion sequences, we propose an \textbf{fast out-painting-based partial autoregressive sampling (FOPPAS)} strategy, which generates partially overlapping clips one by one with a smooth motion at the intersection. Additionally, FOPPAS also features its real-time streaming inference ability. 
In the following sections, we will introduce the problem formulation and diffusion model~\cite{ddpm}, and then elaborate on the design of our proposed framework.

\subsection{Problem Formulation}
Given an arbitrary-long (streaming) audio, we aim to generate realistic, synchronous, and diverse gestures as well as expressions. 
To get training samples, we cut the aligned audio and motion sequences into clips with the same time duration using sliding windows. For each $N$-frame clip, we encode the corresponding audio clip to audio features $\mathbf{A} = [{\mathbf{a}_1, \dots, \mathbf{a}_N}]$. Gesture clip $\mathbf{G} = [{\mathbf{g}_1, \dots, \mathbf{g}_N}]$ is represented by joint rotations in axis-angle format, where $\mathbf{g}_i\in \mathbb{R}^{3J}$ and $J$ is the total joint number. Expression clip $\mathbf{E} = [{\mathbf{e}_1, \dots, \mathbf{e}_N}]$  is represented by the blend shape weights where $\mathbf{e}_i\in \mathbb{R}^{C_{exp}}$ and $C_{exp}$ is the channel dimension of expression. The holistic motion sequence $\mathbf{M}=Concat(\mathbf{G}, \mathbf{E})$, where each motion frame $\mathbf{m}_i\in\mathbb{R}^{3J+C_{exp}}$. 
The training objective of our framework is to reconstruct the motion clip $\mathbf{M}$ conditioned on audio clip $\mathbf{A}$, which is similar to the Multi-Input Multi-Output (MIMO) setting~\cite{bsvdmm}. 
At inference time, we generate realistic, diverse, and well-aligned co-speech expressions and gestures with motion clips smoothly connected.



\subsection{Preliminary: Diffusion Models}
DiffSHEG uses diffusion models~\cite{ddpm} for this task which consists of a diffusion and a denoising process. For explicitness, we substitute notations of expression, gesture, and motion clip  $\mathbf{E},\mathbf{G},\mathbf{M}$ with $\mathbf{x}_0^E, \mathbf{x}_0^G, \mathbf{x}_0$. 
Given a motion clip distribution $p(\mathbf{x}_0)\in\mathbb{R}^{N\times(3J+C_{exp})}$, our goal is to train a model parameterized by $\theta$ to approximate $p(\mathbf{x}_0)$. 

\noindent \textbf{Diffusion Process.} In the diffusion process, models progressively corrupts input data  $\mathbf{x}_0 \sim p(\mathbf{x}_0)$ according to a predefined schedule $\beta_t \in (0, 1)$, eventually turning data distribution into an isotropic Gaussian in $T$ steps. Each diffusion transition can be assumed as
\begin{align}
q\left(\mathbf{x}_{t} \mid \mathbf{x}_{t-1}\right) & = \mathcal{N}\left(\mathbf{x}_{t} ; \sqrt{1-\beta_{t}} \mathbf{x}_{t-1}, \beta_{t} \mathbf{I}\right),
\end{align}
where the full diffusion process can be written as
\begin{align}
q\left(\mathbf{x}_{1: T} \mid \mathbf{x}_{0}\right) & = \prod_{1 \leq t \leq T} q\left(\mathbf{x}_{t} \mid \mathbf{x}_{t-1}\right).
\end{align}
\noindent \textbf{Denoising Process.} 
In the denoising process, models learn to invert the diffusion procedure so that it can turn random noise into real data distribution at inference.
The corresponding denoising process can be written as
\begin{eqnarray}
& p_{\theta}\left(\mathbf{x}_{t-1} \mid \mathbf{x}_{t}\right) = \mathcal{N}\left(\mathbf{x}_{t-1} ; \mu_{\theta}\left(\mathbf{x}_{t}, t\right), \Sigma_{\theta}\left(\mathbf{x}_{t}, t\right)\right) \nonumber\\ &= \mathcal{N}\left(\mathbf{x}_{t-1} ; \frac{1}{\sqrt{\alpha_{t}}}\left(\mathbf{x}_{t}-\frac{\beta_{t}}{\sqrt{1-\bar{\alpha}_{t}}} \mathbf{\epsilon} \right),
\frac{1-\bar{\alpha}_{t-1}}{1-\bar{\alpha}_{t}} \beta_{t}\right),
\end{eqnarray}
where $\mathbf{\epsilon} \sim \mathcal{N}(\mathbf{0},\mathbf{I})$, $\alpha_{t}=1-\beta_{t}$, $\bar{\alpha}_{t}=\prod_{i=1}^{t} \alpha_{i}$ and $\theta$ specifically denotes parameters of a neural network learning to denoise. The training objective is to maximize the likelihood of observed data $p_{\theta}\left(\mathbf{x}_{0}\right)=\int p_{\theta}\left(\mathbf{x}_{0: T}\right) d \mathbf{x}_{1: T}$, by maximizing its evidence lower bound (ELBO), which effectively matches the true denoising model $q\left(\mathbf{x}_{t-1} \mid \mathbf{x}_{t}\right)$ with the parameterized $p_{\theta}\left(\mathbf{x}_{t-1} \mid \mathbf{x}_{t}\right)$. During training, the target of the denoising network $\mathbf{\epsilon}_\theta(.)$ is to restore $\mathbf{x}_0$ given any noised input $\mathbf{x}_t$, by predicting the added noise $\epsilon \sim \mathcal{N}(\mathbf{0},\mathbf{I})$ via minimizing the noise prediction error 
\begin{align}
\mathcal{L}_{t} & = \mathbb{E}_{\mathbf{x}_{0}, \epsilon }\left[\left\|\mathbf{\epsilon}-\mathbf{\epsilon}_{\theta}\left(\sqrt{\bar{\alpha}_{t}}\mathbf{x}_0+\sqrt{1-\bar{\alpha}_{t}} \mathbf{\epsilon}, t\right)\right\|^{2}\right]. \label{equ:noise_loss}
\end{align}
To make the model conditioned on extra context information $\mathbf{c}$, \textit{e.g.}, audio, we inject $\mathbf{c}$ into $\mathbf{\epsilon}_\theta(.)$ by replacing 
$\mu_{\theta}\left(\mathbf{x}_{t}, t\right)$ and $\Sigma_{\theta}\left(\mathbf{x}_{t}, t\right)$ with $\mu_{\theta}\left(\mathbf{x}_{t}, t, \mathbf{c}\right)$ and $\Sigma_{\theta}\left(\mathbf{x}_{t}, t,\mathbf{c}\right)$. 


\subsection{DiffSHEG Framework}
\label{sec:method_resmlp}
Our DiffSHEG consists of audio encoders and the UniEG Transformer generator.
Our UniEG has double meanings: unified joint expression-gesture generation with uni-directional expression-to-gesture condition flow. The UniEG generator includes two main building blocks: Motion-Speech Fusion Residual Block and Style-aware Transformer Block.

\noindent \textbf{Speech Encoding.} 
We convert the raw speech audio into two types of features: Mel-spectrogram and HuBERT~\cite{hsu2021hubert} feature, which serve as low-level and high-level speech features, respectively. The HuBERT encoder is frozen during training. For the low-level Mel-spectrogram feature, we further feed it into a Transformer encoder shared by expression and gesture branch to extract the shared mid-level speech feature. This is drawn upon a classical design of multi-task learning, in which task-specific decoders can share the low-level feature to benefit each other.

\noindent \textbf{Motion-Speech Fusion Residual Block.} 
In order to condition the motion on speech and align them in the temporal dimension, instead of using cross-attention between speech and motion~\cite{fan2022faceformer}, we directly concatenate the motion features with the speech embeddings as well as other
optional temporal conditions along the channel dimension, resulting in a natural temporal alignment between audio and motion as well as an exemption on the use of attention masks in cross-attention temporal conditioning~\cite{fan2022faceformer}.
For feature fusion, instead of using simple linear layer~\cite{zhu2023taming} which leads to slow convergence and unstable training, we design the motion-speech fusion residual block, which fuses concatenated features and projects them into the same shape as motion features as shown in Figure~\ref{fig:framework}. The block consists of a LayerNorm (LN) and an MLP, where the LN ensures numerical stability, and the MLP is designed to predict the residual of the original motion features to facilitate fast convergence (Section~\ref{sec:supp_ResMLP} in Appendix). 

\noindent \textbf{Style-aware Transformer Block.}  
This block injects global conditions (style and diffusion step $t$) to the fused features from the previous Motion-Speech Block. Firstly, style (person ID in this paper) and diffusion step $t$ are projected to vectors of the same shape by MLPs. Then, the summation of two global condition vectors will be fed into AdaIN~\cite{Huang2017AdaIN}.
Two AdaIN stylization blocks are inserted after self-attention and MLP in a Transformer encoder block. The AdaIN blocks apply the feature statistics substitution according to the new statistics computed from the fused global condition vector. Note that we utilize linear self-attention~\cite{shen2021efficient, zhang2022motiondiffuse} as an alternative for full self-attention to save computation and facilitate fast inference.

\noindent \textbf{Uni-directional Expression-to-Gesture Information Flow.} To capture the joint distribution of expression and gesture and facilitate realism and coherence of two types of motion, we feed the predicted expression $\hat{x}_{0(t)}^E$ at the diffusion step $t$ into the gesture Transformer block, as shown in Figure~\ref{fig:framework}. This expression is computed from the predicted expression noise $\hat\epsilon ^ E_t$ at diffusion step $t$:
\begin{equation}
\hat{\mathbf{x}}_{0(t)}^E = \frac{ \mathbf{x}_t^E - \sqrt{1-\bar{\alpha}_t} \hat{\epsilon}^E_t} {\sqrt{\bar{\alpha}_t}}.
\label{eq:x0}
\end{equation}
Note that to prevent the gradient of the gesture branch from affecting the expression encoder, we cut off the gradient of the predicted expression $\hat{x}_{0(t)}^E$ before passing it to the gesture encoder, as illustrated in Figure~\ref{fig:framework}.
The Uni-EG module will repeat $n$ times to get the final noise in each forward of the network. 

\begin{figure}[t]
\begin{center}
\includegraphics[width=0.99\linewidth]{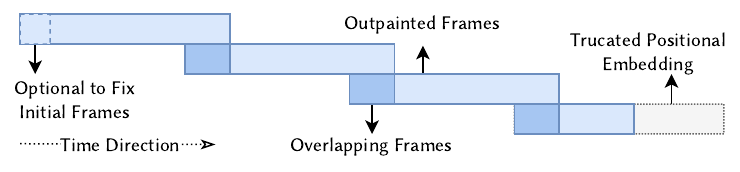}
\vspace{-0.8cm}
\end{center}
   \caption{\textbf{Illustration for outpainting-based arbitrary long sequence inference}. Given a previous clip, we generate current clip by outpainting the remaining frames (light blue) according to the overlaping frames (deep blue). Each row of blue bar represents a motion clip from a single sampling process.}
\label{fig:inpainting}
\vspace{-0.4cm}
\end{figure}

\subsection{Training}
\vspace{-0.15cm}


\noindent \textbf{Loss functions.} Except for the noise prediction loss in Equation~\ref{equ:noise_loss}, we also introduce the velocity loss $\mathcal{L}_{v}$ and Huber motion reconstruction loss $\mathcal{L}_{\delta}$. Since the two losses are computed on the motion $\hat{\mathbf{x}}_0$, we first compute the predicted motion $\hat{\mathbf{x}}_{0(t)}$ from the predicted noise $\hat{\mathbf{\epsilon}_t}$, which is similar to Equation~\ref{eq:x0}.
Then, we have velocity loss computed as the mean square error of the velocity:
\begin{equation}
\mathcal{L}_{v} = \mathbb{E}\left[\left\|(\mathbf{x}_0[1:] - \mathbf{x}_0[:-1]) - (\hat{\mathbf{x}}_0[1:] - \hat{\mathbf{x}}_0[:-1]) \right\|^{2}\right], 
\end{equation}
and the Huber loss for the reconstruction of motion:
\begin{equation}
\mathcal{L}_{\delta}=
    \begin{cases}
        \frac{1}{2}(\mathbf{x}_0 - \hat{\mathbf{x}}_0)^{2}, & \text{if} \left | (\mathbf{x}_0 - \hat{\mathbf{x}}_0)  \right | < \delta , \\
        \delta ((\mathbf{x}_0 - \hat{\mathbf{x}}_0) - \frac1 2 \delta), & \text{otherwise.}
    \end{cases}
\end{equation}
The final loss is a weighted sum of the three losses:
\begin{equation}
\mathcal{L} = \lambda_t \mathcal{L}_t + \lambda_v \mathcal{L}_v + \lambda_\delta \mathcal{L}_\delta, 
\end{equation}
where $\lambda_t=10$, $\lambda_v=1$, and $\lambda_\delta=1$ in our experiment.

\subsection{Arbitrary-long Motion Generation}\label{sec:inferene}

Instead of conditioning the model on previous frames during training~\cite{ijcai2023p650, zhu2023taming, beat}, we propose to realize the arbitrary long sampling via outpainting at the test time without training, which has more flexibility and less computation waste. 

\noindent \textbf{Fast Outpainting-based Partial Autoregressive Sampling (FOPPAS)}. As illustrated in Figure~\ref{fig:inpainting}, starting from the second clip, we can fix the initial frames to be the same as the last frames of the previous clip and then outpaint the remaining frames in this clip. This method conditions the current clip only on the part of the previous clip frames; therefore, we call it partial autoregressive sampling via outpainting. Unlike the RNN-based methods~\cite{beat} or train-time autoregressive diffusion models~\cite{zhu2023taming, ijcai2023p650}, the number of overlapping frames of FOPPAS can be flexibly set and changeable anytime instead of being required and fixed once trained. Those methods~\cite{beat, zhu2023taming, ijcai2023p650} also require an initial "seed motion" to generate the first clip. However, it is not convenient for users to find such seed sequences or they just want to generate the first clip randomly. In contrast, we can generate the first clip with the overlapping number set as 0. In our experiment, we choose Repaint~\cite{lugmayr2022repaint} to perform diffusion-based outpainting.


\noindent \textbf{Shorter Clip Sampling.} When using Transformer without positional embedding, it becomes a point-wise set function. Therefore, the length of an ordered sequence that a Transformer can process in one pass only depends on the length of the positional embedding used during training. Thanks to this property of the Transformer, our framework can infer any clip that is shorter than the training clip, which is achieved by dropping the remaining positional embedding. This is particularly useful when inferring the last clip, as demonstrated in Figure~\ref{fig:inpainting}.

\noindent \textbf{Towards Real-time Sampling.}
The original DDPM has 1000 steps to generate a clip, which is very slow in practice. Moreover, the total denoising steps would be further increased due to the re-sampling operation in Repaint~\cite{lugmayr2022repaint}.
To enable real-time generation for various applications in the development of digital humans and embodied agents, we adapt the Repaint~\cite{lugmayr2022repaint} algorithm to DDIM~\cite{ddim} sampling. We use 25-step DDIM to replace the 1000-step DDPM, with a speedup of about 40 times during inference.

\noindent \textbf{Last-steps Refinement with Blending.} Although applying outpainting already gets a very smooth transition between clips, we occasionally observe slight inconsistency at the boundary. To refine and facilitate the consistency at the clip boundary, we perform a linear blending at the overlapping part of clips at the last two sampling steps. For the details of FOPPAS, please refer to Algorithm~\ref{alg:ddim_repaint} in the Appendix.

\begin{figure*}[t]
\begin{center}
\includegraphics[width=0.92\linewidth]{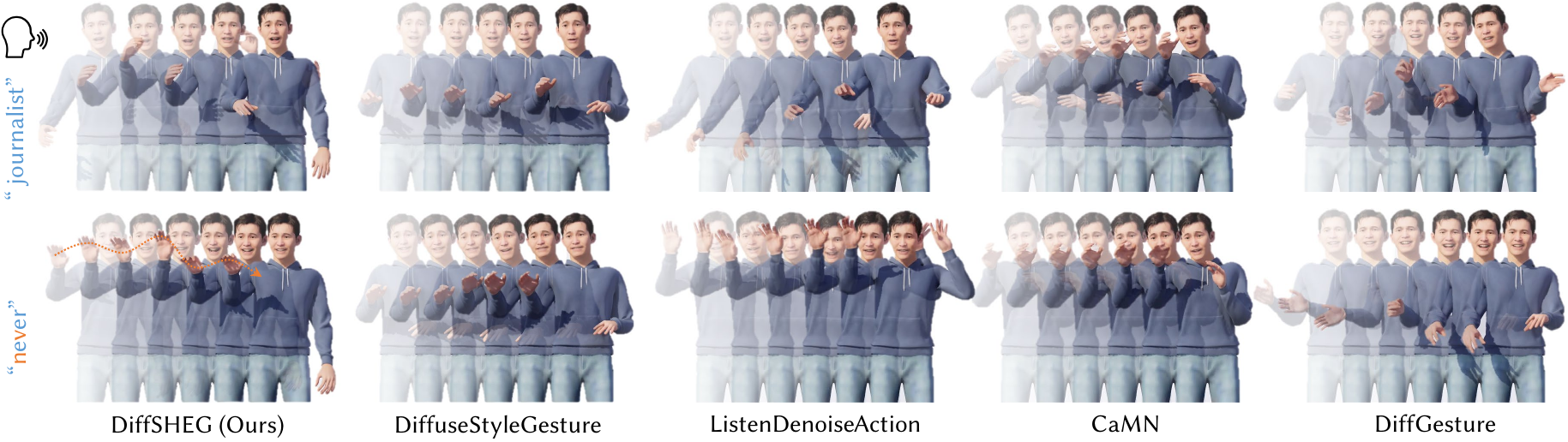}
\vspace{-0.5cm}
\end{center}
   \caption{\textbf{Qualitative Comparison on BEAT~\cite{{beat}} Dataset.}  In comparison to baseline methods, our approach generates a broader range of natural, agile, and diverse gestures that are closely synchronized with the audio input. When saying "journalist", the character driven by our motion raises double hands to stress this word; When saying "never", our motion shows two times up-and-down right hand and fingers, corresponding to the two syllables ``ne" and ``ver". The character is from MetaHuman~\cite{metahuman} rendered by Unreal Engine 5~\cite{ue5}.
   }
\label{fig:beat_comparison}
\vspace{-0.40cm}
\end{figure*}

\begin{figure*}[t]
\begin{center}
\includegraphics[width=0.92\linewidth]{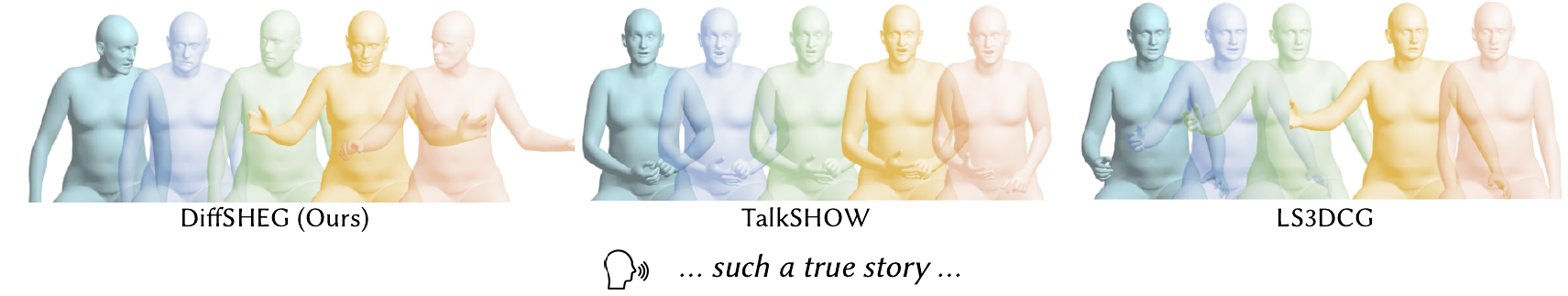}
\vspace{-0.7cm}
\end{center}
   \caption{\textbf{Motion Comparison on the SHOW~\cite{talkshow} Dataset.} Our method generates more expressive and diverse motions than TalkShow~\cite{talkshow} and LS3DCG~\cite{habibie2021learning} in terms of both gesture and head pose diversity. Our results also show more agile motions than baselines.  }
\label{fig:motion_dieversity}
\vspace{-0.4cm}
\end{figure*}

\begin{figure}[t]
\begin{center}
\includegraphics[width=0.99\linewidth]{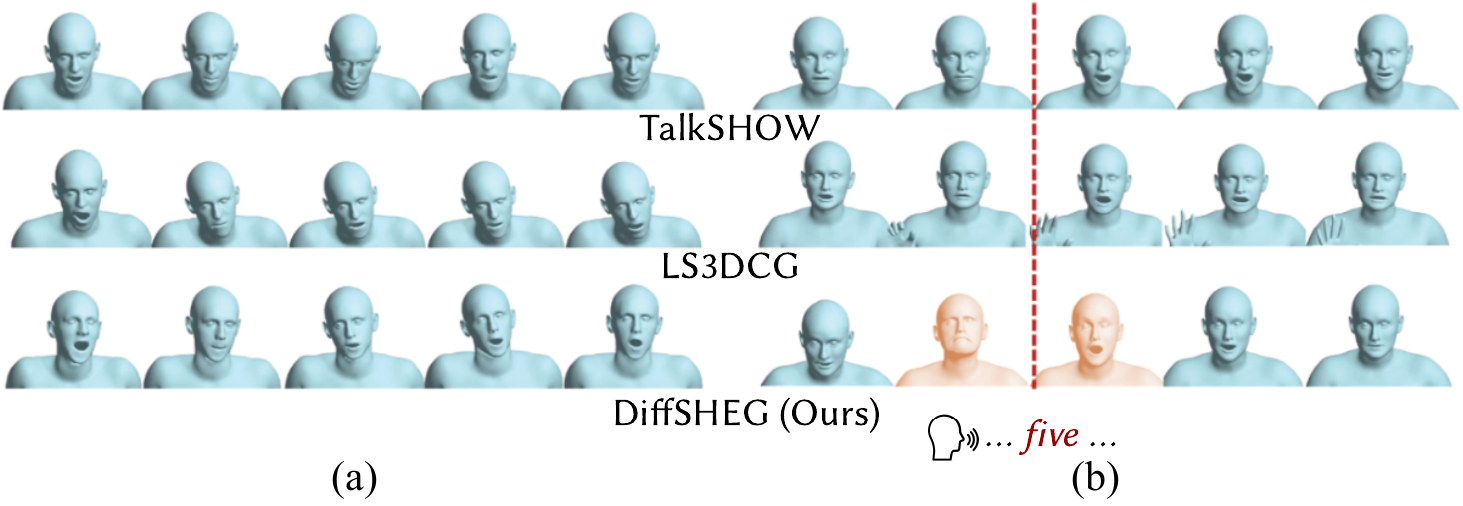}
\vspace{-0.6cm}
\end{center}
   \caption{\textbf{Expression and Head Pose Comparison on the SHOW~\cite{talkshow} Dataset.}
   (a) With speech audio as input, TalkShow~\cite{talkshow} and LS3DCG~\cite{habibie2021learning} may generate unnatural and persistent head-down poses showing limited variation. In contrast, our method produces a wide range of expressive head poses. (b) Prior to the audio input's emphasis on the word ``five'', our approach instinctively raises the head to prepare for highlighting, subsequently producing precise lip movements accompanied by raised eyebrows to emphasize the word ``five''.
   }
\label{fig:head_and_expression}
\vspace{-0.7cm}
\end{figure}

\vspace{-0.2cm}
\section{Experiments}
\label{sec:experiment}

\subsection{Datasets}
\vspace{-0.15cm}
\textbf{BEAT}~\cite{beat} is a large-scale, multi-modal human gestures and expressions dataset with text, semantic and emotional annotations. We follow the train-validation-test split setting in~\cite{beat} with four subjects. The training and validation samples are 34-frame clips. The test samples have 64 long sequences with a duration of around 1 minute. The motions are then resampled into 15 frames per second.
We adopt axis-angle rotation representation for smooth motion generation. 
\textbf{SHOW~\cite{talkshow}} is a new audio-visual dataset that includes SMPLX~\cite{SMPL-X:2019} parameters of 4 persons reconstructed from videos at 30 fps, along with corresponding synchronized audio sampled at 22K rate. Following the same settings in SHOW~\cite{talkshow}, we generate the SMPLX parameters according to the audio. 
The training and validation samples are 88-frame motion clips, and the test samples are long motion sequences with different lengths.

\subsection{Experimental Setup}
\vspace{-0.15cm}
We train our model on five 3090 GPUs. For \textbf{BEAT}, we train our model for 1000 epochs with a batch size of 2500. For \textbf{SHOW}, we train 1600 epochs with a batch size of 950.

\noindent \textbf{Baselines.}
We compare our method with CaMN~\cite{beat}, DiffGesture~\cite{zhu2023taming}, DiffuseStyleGesture~\cite{ijcai2023p650} and ListentDenoiseAction (LDA)~\cite{alexanderson2023listen} on BEAT. CaMN is proposed with the BEAT dataset, which is based on LSTM and can fuse multiple 
conditions, including audio, text, face, and emotion. DiffGesture, DiffuseStyleGesture, and LDA are newly proposed diffusion-based co-speech generation methods. 
For gesture generation, the four baselines are retrained on the axis-angle rotation representation instead of the Euler angles.
For expression generation, since four baseline methods were originally proposed for gesture generation solely, we substitute the gesture into the expression data to generate expressions independently. For a fair comparison, we condition all the models on audio and person ID.
On the SHOW dataset, we utilize two baselines that focus on joint expression and gesture generation: \textbf{TalkSHOW}~\cite{talkshow} and the method proposed in~\cite{habibie2021learning} named \textbf{LS3DCG} here. 
Since all the baselines only focus on the upper body movement for gesture generation, we follow this setting despite that our framework also works for the lower body motion.


\begin{figure*}[t]
\begin{center}
\includegraphics[width=0.99\textwidth]{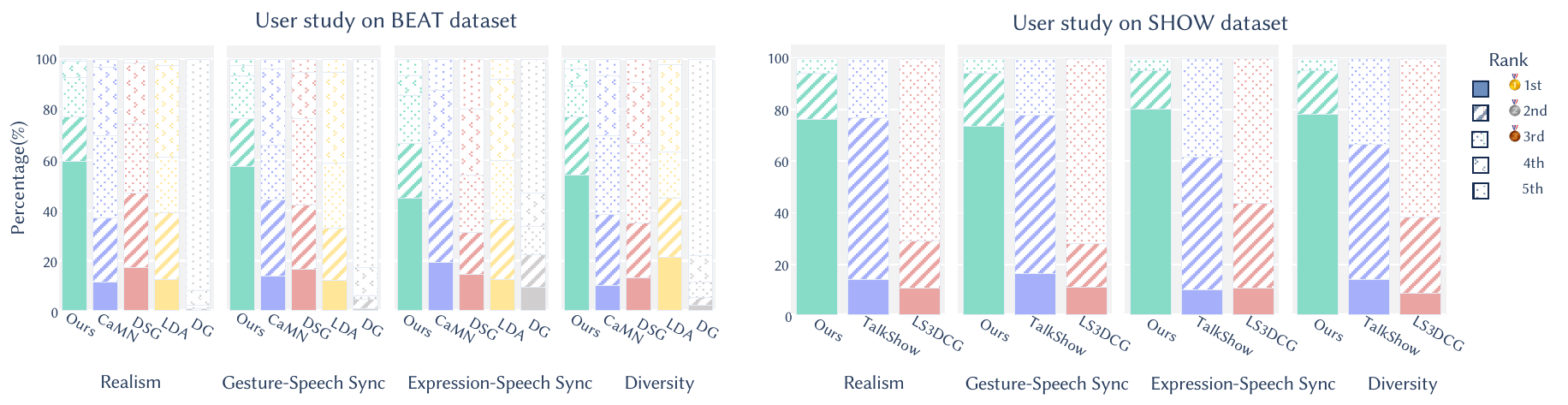}
\vspace{-0.7cm}
\end{center}
   \caption{\textbf{Results of the user study}. The chart shows user preference percentage in terms of four metrics: realism, gesture-speech synchronism,
   expression-speech synchronism, and motion diversity. In both datasets and all metrics, our method is dominantly preferred. DSG and DG are the abbreviations of DiffuseStyleGesture and DiffuseGesture.}
\label{fig:user_study}
\vspace{-0.15cm}
\end{figure*}

\begin{table*}[t]\footnotesize
    \centering
    
    \newcolumntype{Z}{>{\centering\arraybackslash}X}
    \begin{tabularx}{\linewidth}{clZZZZZZZZZ}
        \toprule
        \multirow{2}{*}{Dataset} & \multirow{2}{*}{Method} & \multicolumn{1}{c}{{Holistic}} & \multicolumn{3}{c}{{Expression}} & \multicolumn{5}{c}{{Gesture}} \\ 
        \cmidrule(lr){3-3}\cmidrule(lr){4-6}\cmidrule(lr){7-11}
    
        \multicolumn{2}{c}{{}} & \multicolumn{1}{c}{{ FMD $\downarrow$}} & \multicolumn{1}{c}{{ FED $\downarrow$ }}  & \multicolumn{1}{c}{{Div $\uparrow$ }} & \multicolumn{1}{c}{{Gen}} & \multicolumn{1}{c}{{FGD $\downarrow$}}
        & \multicolumn{1}{c}{{BA$\uparrow$ }} & \multicolumn{1}{c}{{SRGR $\uparrow$}} & \multicolumn{1}{c}{{Div $\uparrow$ }}& \multicolumn{1}{c}{{Gen}}\\
     
        \toprule
        \multirow{9}{*}{BEAT~\cite{beat}} & \groundtruth{Ground Truth} & \groundtruth{-} & \groundtruth{-}  & \groundtruth{0.651} & \groundtruth{-} & \groundtruth{-} & \groundtruth{0.915} & \groundtruth{0.961} & \groundtruth{0.819} & \groundtruth{-}  \\
        \cline{2-11}
        & CaMN (ECCV'22)~\cite{beat}         & 1055.52 & 1324.00 & 0.479 & \XSolidBrush & 1635.44 & 0.793 & 0.197 & 0.633 & \XSolidBrush \\
        & DiffGesture (CVPR'23)~\cite{zhu2023taming}     & 12142.70 & 586.45 & 0.625 & \Checkmark &  23700.91 & 0.929 & 0.096 & 3.284 & \Checkmark \\
        & DSG (IJCAI'23)~\cite{ijcai2023p650}     & 1261.59 & 998.25 & 0.688 & \Checkmark &  1907.58 & 0.919 & 0.204 & 0.701 & \Checkmark \\
        & LDA (SIGGRAPH'23)~\cite{alexanderson2023listen}     & 688.25 & 510.345 & 0.603 & \Checkmark &  997.62 & 0.923 & 0.215 & 0.688 & \Checkmark \\
        \cline{2-11}
        & Ours Na\"ive Concat    & 354.60 & 354.60 & 0.526 & & 497.28 & 0.914 & 0.253& 0.503 \\
        & Ours w/o Detach & 375.98 & 384.31 & 0.513 & & 475.19 & 0.917 & 0.247 & 0.517 \\
        & Ours w/o $\hat{x}_{0(t)}^E$  & 369.97 & 366.37 & 0.519 & \Checkmark & 477.00 & 0.917 & 0.251 & 0.504 & \Checkmark\\
        & Ours Reverse Direction  & 357.56 & 369.72 & 0.583 & & 472.38 & 0.913 & 0.245 & 0.553 \\
        & \textbf{DiffSHEG (Ours)}            & \textbf{324.67} & \textbf{331.72} & 0.539 & & \textbf{438.93} & 0.914 & 0.251 & 0.536 \\

        \midrule
        \multirow{8}*{SHOW~\cite{talkshow}} & 
        \groundtruth{Ground Truth} & \groundtruth{-} & \groundtruth{-} &  \groundtruth{0.990} & \groundtruth{-} & \groundtruth{-} & \groundtruth{0.867} & \groundtruth{PCM=1} & \groundtruth{0.834} & \groundtruth{-}\\        
        \cline{2-11}
        & LS3DCG* (IVA'21)~\cite{habibie2021learning}     & 0.00230 & 0.00229 & 0.708 & \XSolidBrush & 0.00478 & 0.947 & 0.981 & 0.645 & \XSolidBrush  \\
        & TalkSHOW* (CVPR'23)~\cite{talkshow}   & 0.00219 & 0.00233 & 0.740 & \XSolidBrush & 0.00323 & 0.869 & 0.902 & 0.703 & \Checkmark \\
        & TalkSHOW (Re-train)~\cite{talkshow} & 0.00278 & 0.00408 & 0.618 & \XSolidBrush & 0.00328 & 0.872 & 0.894 & 0.711 & \Checkmark \\
        \cline{2-11}
        & Ours Na\"ive Concat   &  0.00255 & 0.00192 & 0.766 & & 0.00407 & 0.896 & 0.932 & 0.618 & \\
        & Ours w/o Detach & 0.00259 & 0.00234 & 0.745 & & 0.00378 & 0.899 & 0.926 & 0.673 & \\
        & Ours w/o $\hat{x}_{0(t)}^E$  & 0.00248 & 0.00251 & 0.728 & \Checkmark & 0.00311 & 0.892 & 0.927 & 0.694 & \Checkmark \\
        & Ours Reverse Direction  & 0.00269 & 0.00266 & 0.731 &  & 0.00347 & 0.882 & 0.926 &  0.695 &  \\
        & \textbf{DiffSHEG (Ours)}       & \textbf{0.00184} & \textbf{0.00161} &  0.923 & & \textbf{0.00271} & 0.902 & 0.912 & 0.810 & \\
        \bottomrule 
    \end{tabularx}
    \vspace{-0.2cm}
    \caption{\textbf{Quantitative comparison and ablation study}. On the BEAT~\cite{beat} dataset, we compare our DiffSHEG with CaMN~\cite{beat}, DiffGesture~\cite{zhu2023taming}, DiffuseStyleGesture (DSG)~\cite{ijcai2023p650} and LDA~\cite{alexanderson2023listen} with audio and person ID as input. Note that the baseline methods are originally for gesture generation solely, and we apply the same procedure independently for expression generation. On the SHOW~\cite{talkshow} dataset, we compare with LS3DCG~\cite{habibie2021learning} and TalkSHOW~\cite{talkshow}. The ablation studies are conducted on both datasets to demonstrate the effectiveness of our UniEG-Transformer design. Note that we use SRGR on the BEAT dataset and PCM on SHOW dataset. *: indicates that the results are computed using the pre-trained checkpoints provided by authors of TalkSHOW~\cite{talkshow}. }
    \label{tab:table1}
    \vspace{-0.26cm}
\end{table*}

\subsection{Qualitative Results}
\vspace{-0.14cm}

\noindent \textbf{Qualitative comparisons.} We highly recommend readers watch our demo video to get an intuitive sense of qualitative results. Overall, in addition to better speech-motion alignment, we observe more \textbf{realistic}, \textbf{agile} and \textbf{diverse} expressions and gestures of ours than that of baselines on both datasets 
(Figure~\ref{fig:beat_comparison},~\ref{fig:motion_dieversity},~\ref{fig:head_and_expression}). 
On the BEAT dataset, our method shows more agile gestures than CaMN. The gestures of baseline diffusion-based methods all suffer different extents of jittering. The jittering extents from high to low are DiffGesture, DiffuseStyleGesture, and LDA. Both our method and CaMN can generate smooth motions. However, CaMN exhibits much slower and less diverse motion while ours is as agile as the real motion.
For expression, CaMN shows inaccurate lip sync while other methods all look fine.
Interestingly, we find that the CaMN barely blinks, while our method can generate good expressions with a reasonable frequency of blinks. Other baselines show a more frequent blinking, corresponding to higher diversity scores. However, ours has the most similar blink frequency as the real data.
On the SHOW dataset, our method shows more agile gestures than TalkSHOW and more diverse expressions and gestures (Figure~\ref{fig:motion_dieversity},~\ref{fig:head_and_expression}). This might be due to the fact that Talkshow utilizes VQ-VAE to encode the gestures, which may filter the high-frequency gestures, leading to slow motion. Also, because VQ-VAE tokenizes the gestures into limited numbers, its gestures show lower diversity than ours. The LS3DCG suffers from jittering arms, which may be because their deterministic CNN structure has difficulty approximating the many-to-many mapping. Note that we retrain the TalkSHOW for qualitative comparison because the results of TalkSHOW and LS3DCG with author-provided checkpoints have many highly similar concurrent motions and face distortions as ground truth during testing.

\noindent \textbf{User study.}
We conducted a user study on eight 1-minute-long videos in BEAT and twelve 10-second-long videos in SHOW sampled from the test set. We recruited 22 subjects with diverse backgrounds to evaluate \textbf{holistic realism}, \textbf{expression-speech synchronism}, \textbf{gesture-speech synchronism}, and \textbf{holistic diversity} given real data as a reference. Users are instructed to perform sorting questions on shuffled videos of different methods. Note that for diversity, we let users sort the motion diversity under the premise of smooth and natural motion. The result is reported in Figure~\ref{fig:user_study}. 
The user study results show that our generated expressions and motions are dominantly preferred on all four metrics over the baseline method, demonstrating that our proposed DiffSHEG approach is capable of generating more realistic, synchronized, and diverse expressions and gestures that humans prefer. The preference for our generated motions on realism also demonstrates the effectiveness of our out-painting-based sampling strategy, which has a smooth connection between adjacent clips.

\vspace{-0.10cm}
\subsection{Quantitative Results}
\vspace{-1.2pt}
\textbf{Metrics.} Except for the gold standard of evaluation for generative tasks --- user study, we also provide several quantitative metrics as a reference: 
(1) \textbf{FMD}, \textbf{FGD}, and \textbf{FED} denotes Fr\'echet Motion Distance, Fr\'echet Gesture Distance, and Fr\'echet Expression Distance. Except for the FGD~\cite{yoon2020speech} that is proposed to evaluate the Fr\'echet distance between generated and real data in gesture feature space.
Similarly, we propose FMD and FED that can indicate the generation-GT distribution distance for holistic motion (joint expression-gesture) distribution and expression distribution.
(2) \textbf{PCM} and \textbf{SRGR}. PCM is the Percent of Correct Motion parameters,
which is computed on motion parameters instead of keypoints in our experiment. SRGR is a weighted version of PCM according to the temporal semantic weight proposed in BEAT~\cite{beat}. We use PCM for SHOW and SRGR for BEAT.
(3) \textbf{Diversity (Div)} is used to evaluate whether a model can generate a wide range of dynamic and varied motions, following~\cite{bhattacharya2021speech2affectivegestures}. This metric involves calculating the distance between various generated gestures. 
(4)~Beat alignment (\textbf{BA})~\cite{Li_2021_aist} is a Chamfer Distance between audio and gesture beats to evaluate gesture-audio beat similarity and we follow the same implementation in the BEAT paper~\cite{beat}.
(5) We also mark in Table~\ref{tab:table1} to show whether it is a generative (\textbf{Gen}) model. Generative models can generate different and diverse motions from the same audio input, while the deterministic models output the determined motion given the same input speech.
Note that except for the Fr\'echet distances, other quantitative metrics only serve as a reference since they are not always aligned with the human-perceived visual quality~\cite{alexanderson2023listen,ao2022rhythmic}.

\noindent \textbf{Comparison with Baselines.} 
The results in Table~\ref{tab:table1} demonstrate our method can achieve state-of-the-art performance compared with other baselines on both datasets. We consistently outperform the baseline methods on Fr\'echet distance (\textbf{FMD}, \textbf{FED}, and \textbf{FGD}) by a large margin, indicating the strong distribution matching ability of DiffSHEG, especially the expression-gesture joint distribution. For SRGR and PCM of gestures, we outperform all the baselines except for the LS3DCG which generates jittering motion. 
Note that \textbf{Div} and \textbf{BA} are only meaningful when the synthesized motions are smooth and natural because the jittering motion or motion with outliers can also result in a high score of diversity and beat alignment. For jittering motion, the beat detector will regard each jitter timestep as a beat covering most of the audio beat. 
On the BEAT dataset, we outperform CaMN on the gesture BA and get a comparative BA with other diffusion baselines. As for the Div score, ours is higher than CaMN but lower than other diffusion baselines. This aligns with the qualitative observations that CaMN has slow motions and other diffusion baselines suffer from jittering.
On the SHOW dataset, our DiffSHEG is consistently better than TalkSHOW and LS3DCG on both BA and Div, except that LS3DCG has a higher BA which may be due to the jittering. The Div of our DiffSHEG for both expression and gesture achieves a similar score to that of real data, indicating that the diversity of our generated results can achieve a realistic level on the SHOW dataset.

\noindent \textbf{Impact of FOPPAS and Runtime Analysis.} Diffusion models suffer from the long sampling steps at the test time -- the original DDPM model takes 1000 steps to sample a quality image. Considering the resampling of Repaint~\cite{lugmayr2022repaint}, it would be slower by times.  However, thanks to our FOPPAS mechanism, we can synthesize arbitrary-long smooth motion sequences in real time. We test the runtime of our method on a one-minute-long audio corresponding to 900 frames of expressions and gestures (BEAT uses 15 FPS motion), with the number of overlapping frames in FOPPAS set to 4. The average runtime of our method is 28.6 seconds on a single NVIDIA GeForce RTX 3090 GPU, corresponding to around 31.5 FPS. The time for audio encoding (Mel-Spectrogram and HuBERT) is also included. In contrast, if we directly use DDPM with Repaint~\cite{lugmayr2022repaint}, it costs 2068.1s to infer 900 frames using the default setting of repaint, which is 0.44 FPS. Therefore, our proposed FOPPAS can benefit the diffusion model to achieve real-time in the inference stage, making many downstream products applicable.  


    

    

\vspace{-0.05cm}
\subsection{Ablation Study}
\vspace{-0.1cm}
\label{sec:ablation}
\textbf{Uni-directional condition flow.} To demonstrate the effectiveness of our uni-directional condition flow design, we conduct experiments on three ablated architectures: ours without expression condition flow into gesture encoder (\textbf{Ours w/o $\mathbf{\hat{x}_0^E}$}), ours without cutting off the gradient when feeding predicted expression to gesture encoder (\textbf{Ours w/o Detach}), na\"ively concatenate expression and gesture together to be fed into a single Transformer denoising block (\textbf{Ours Na\"ive Concat}), and reverse the direction of condition flow to become gesture to expression (\textbf{Ours Reverse Direction}). For Ours Na\"ive Concat, we enlarge the latent dimension of the Transformer block by 1.41 times to have a similar computational complexity for a fair comparison. 
From Table~\ref{tab:table1}, on both BEAT and SHOW datasets, our DiffSHEG outperforms all the four ablated architectures on Fr\'echet distances and achieves better or comparable BA, Diversity, and PCM/SRGR scores. On the BEAT dataset, the reverse direction ablation has a higher diversity due to the high entropy of gestures as a condition for expression, but its other metrics are still worse than DiffSHEG, indicating the adverse affect of gesture condition flow to expression. These results demonstrate the effectiveness of our design and validate our intuitive analysis of the relationship between expression and gesture.  
For more ablations, please refer to our Appendix.

\vspace{-0.1cm}
\section{Conclusion}
\vspace{-0.2cm}
In conclusion, we propose a novel diffusion-based approach, DiffSHEG, for speech-driven holistic 3D expression and gesture generation. Our approach addresses the joint expression and gesture generation challenges, which have received less attention in previous works. We develop a Transformer-based framework, UniEG-Transformer, that enables the uni-direction flow of information from expression to gesture in the high-level feature space, which can effectively capture the expression-gesture joint distribution. We also introduce a fast outpainting-based partial autoregressive sampling (FOPPAS) strategy with DDIM to achieve real-time inference on arbitrary-long streaming audio. Our approach has demonstrated state-of-the-art performance on two public datasets quantitatively and qualitatively. The qualitative results and user study show that our method can generate realistic, agile, and diverse expressions and gestures that are well-aligned with speech.

\vspace{-0.1cm}
\section{Acknowledgement}
\vspace{-0.2cm}
We sincerely thank the help from Shaojun Liu and Zhaoxin Fan on the MetaHuman character rendering.

{   
    \clearpage
    \small
    \bibliographystyle{ieeenat_fullname}
    \bibliography{refbib.bib}
}

\clearpage
\setcounter{page}{1}
\maketitlesupplementary


\appendix

\label{sec:supp}

\section{User Study}
\begin{figure}[t]
\begin{center}
\includegraphics[width=0.99\linewidth]{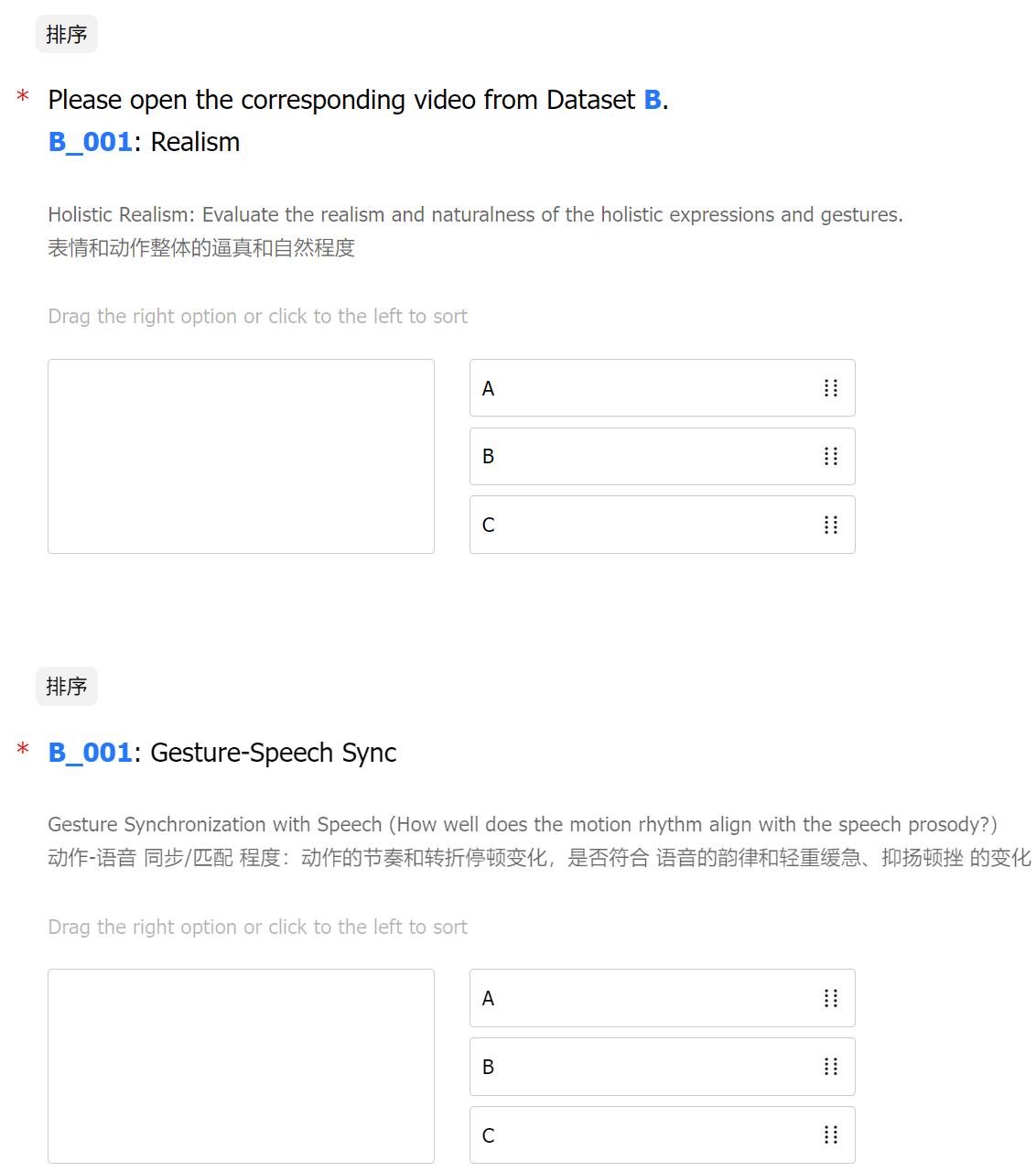}
\end{center}
   \caption{Part of the questionnaire of our user study. We ask users to sort the results of different methods according to the metrics.}
\label{fig:questionnaire}
\end{figure}

\begin{figure}[t]
\begin{center}
\includegraphics[width=0.99\linewidth]{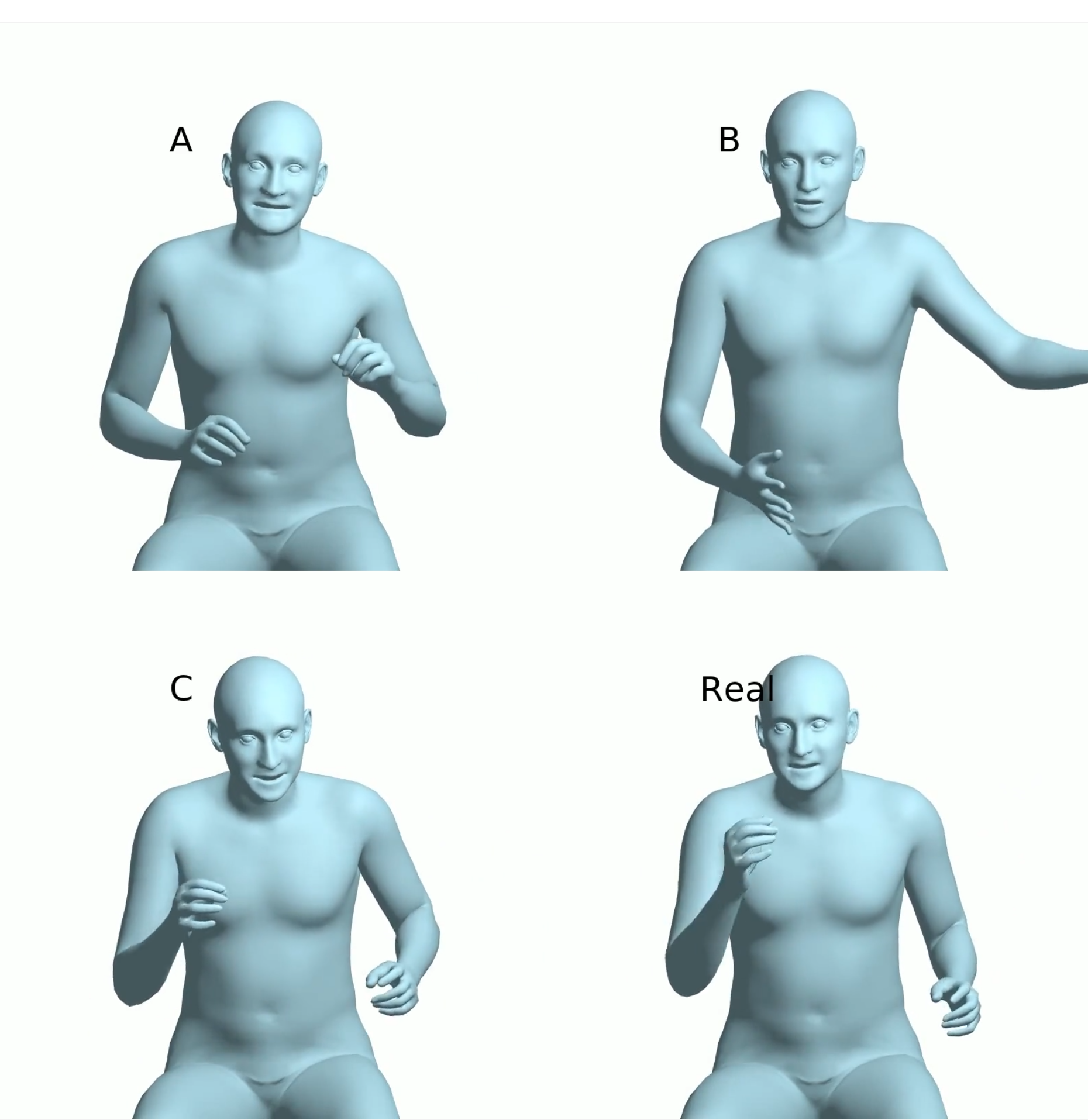}
\end{center}
\vspace{-0.5cm}
   \caption{A screenshot of our user study video.}
\label{fig:video}
\end{figure}

We utilize the Tencent Questionnaire system to conduct the user study. Figure~\ref{fig:questionnaire} is a screenshot of part of our questionnaire. The users are asked to download the videos to their local disk to avoid potential latency and out-of-synchronization between audio and video. The video order is shuffled to avoid inertia bias. Figure~\ref{fig:video} is a screenshot of our shuffled video. The average time to complete the questionnaire is 40 min. To avoid invalid data, We dropped the questionnaires that took shorter than 10 min.

\section{Algorithm for FOPPAS}
To clearly describe the process of FOPPAS, we provide Algorithm~\ref{alg:ddim_repaint} for reference. Algorithm~\ref{alg:ddim_repaint} describes the single clip pass of our FOPPAS method in Sec.~\ref{sec:inferene}.
Specifically, we utilize Repaint~\cite{lugmayr2022repaint} as our out-painting method and adapt it to DDIM sampling to boost the inference speed.

\begin{algorithm}[t]
\caption{Sampling one clip in FOPPAS} \label{alg:ddim_repaint}
\textbf{HyperParameter}: Resampling Steps $U$, Overlapping Length $L_o$, Clip Length $L$, Diffusion Steps T. \\
\textbf{Input (Optional)}: Overlapping motion $x_0^{\text{known}} \in R^{L_o \times C}$ \\
\textbf{Output}: Generated motion $x_0$ 
\begin{algorithmic}[1]
    \State $x_t \sim \mathcal{N}(\mathbf{0}, \mathbf{I})$\Comment{$x_t \in R^{L \times C}$, where $C$ is channel size.}
    \vspace{1mm}
    \State $m = [1,...,1,0,...,0]$\Comment{1 indicates $L_o$ overlapping frames. The length of $m$ is $L$.}
    \vspace{1mm}
    \State $w = LinearSpace([0,1], num=L_o) $
    \vspace{1mm}
    \State $w = PadZero(w, num=L-L_o) $
    \vspace{1mm}
    \State $x_0^{known} = PadZero(x_0^{\text{known}}, num=L-L_o)$
    \vspace{1mm}
    \For{$t=T, \dotsc, 1$}
        \For{$u=1, \dotsc, U$}
        \vspace{1mm}
          \State $\epsilon \sim \mathcal{N}(\mathbf{0}, \mathbf{I})$ if $t > 1$, else $\epsilon = \mathbf{0}$
          \vspace{1mm}
          \State $x_{t-1}^\text{known} = \sqrt{\bar{\alpha}_{t-1}} x_0 + \sqrt{1-\bar{\alpha}_{t-1}} \epsilon$
          \vspace{1mm}
          \State $ \hat{x}_0^t = 
          \frac{ x_t - \sqrt{1-\bar{\alpha}_t} \mathbf{\epsilon}_\theta(x_t, t)} {\sqrt{\bar{\alpha}_t}}
          $
          \vspace{1mm}
          \State $x_{t-1} =
          \sqrt{\bar\alpha_{t-1}}  
          \hat{x}_0^t + 
          \sqrt{1-\bar{\alpha}_{t-1}} \mathbf{\epsilon}_\theta(x_t, t)
          $ 
          \vspace{2mm}
            \If{$t > 2$}
                \State $x_{t-1} = m \odot x_{t-1}^\text{known} + (1-m) \odot x_{t-1}$
            \Else
                \State $x_{t-1}^\text{blend} = m \odot ((1-w) \odot x_{t-1}^\text{known} + w  \odot x_{t-1})$
                \State $x_{t-1} = m \odot x_{t-1}^\text{blend} + (1-m) \odot x_{t-1}$
                \vspace{1mm}
            \EndIf
            \If{$u < U ~\text{and}~ t > 1$}
              \State $x_t \sim \mathcal{N}(\sqrt{1-\beta_{t-1}} x_{t-1}, \beta_{t-1} \mathbf{I})$
            \EndIf
        \EndFor
    \EndFor
    \State \textbf{return} {$x_0$}
\vspace{.04in}
\end{algorithmic}
\end{algorithm}

\section{Effectiveness of Motion-Speech Residual Block }
\begin{figure}[t]
\begin{center}
\includegraphics[width=1\linewidth]{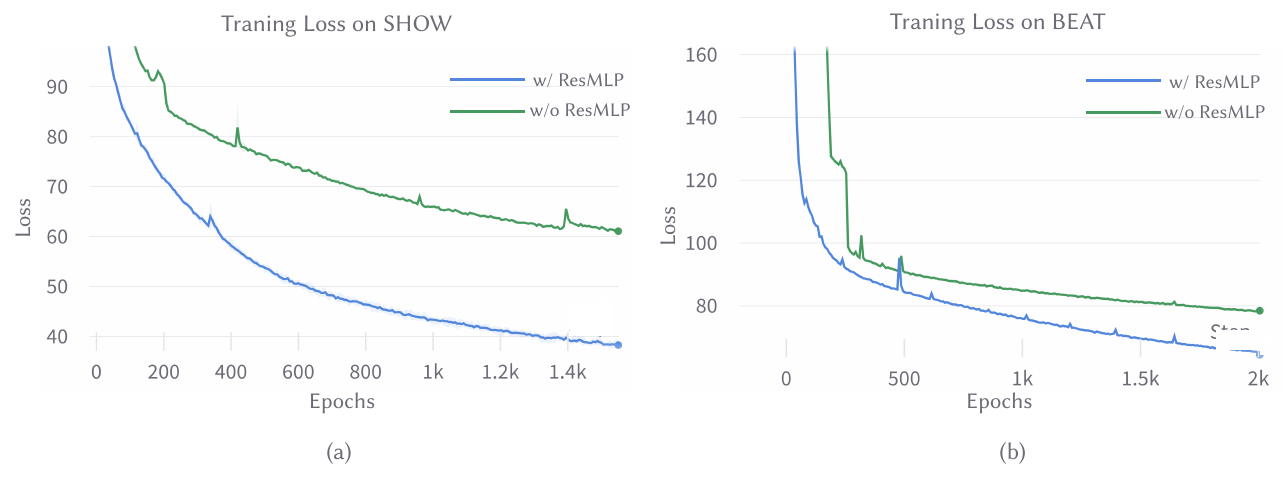}
\end{center}
\vspace{-0.5cm}
   \caption{The training loss of our DiffSHEG on the (a) SHOW and (b) BEAT datasets with and without the motion-speech residual block. We can see that our motion-speech residual block helps the diffusion training process converge much faster.}
\label{fig:resmlp}
\vspace{-0.5cm}
\end{figure}

As shown in Figure~\ref{fig:resmlp}, the utilization of our motion-speech residual block can boost the convergence speed by 2$\sim$4 times. Instead of directly using linear projection~\cite{zhu2023taming} for motion-speech fusion, we utilize an MLP residual block as mentioned in Sec.~\ref{sec:method_resmlp}.

\label{sec:supp_ResMLP}
\section{Analysis on Agility and Smoothness} 
We find our method shows the same level of agility and smoothness as the real motions. In contrast, most baselines fail to match the realistic level of agility and smoothness.

\noindent \textbf{Metrics for agility and smoothness.} As shown in Table~\ref{tab:smoothness}, we compute three metrics for smoothness and agility: AE, Vel, and MLVS, which are the mean acceleration error, mean velocity, and mean local velocity standard deviation of 3D joint positions, respectively. Our DiffSHEG and CaMN have the smallest AE, corresponding to the best smoothness. On the other hand, our DiffSHEG has the closest Vel and MLVS to GT, while CaMN and DSG show very low Vel and MLVS, which is consistent with the observation of their slow motions. Similarly, the jitter motion of LDA and DG leads to high values in three metrics.

\begin{figure*}[t]
\centering
\begin{subfigure}[t]{0.99\textwidth}
   \includegraphics[width=0.99\linewidth]{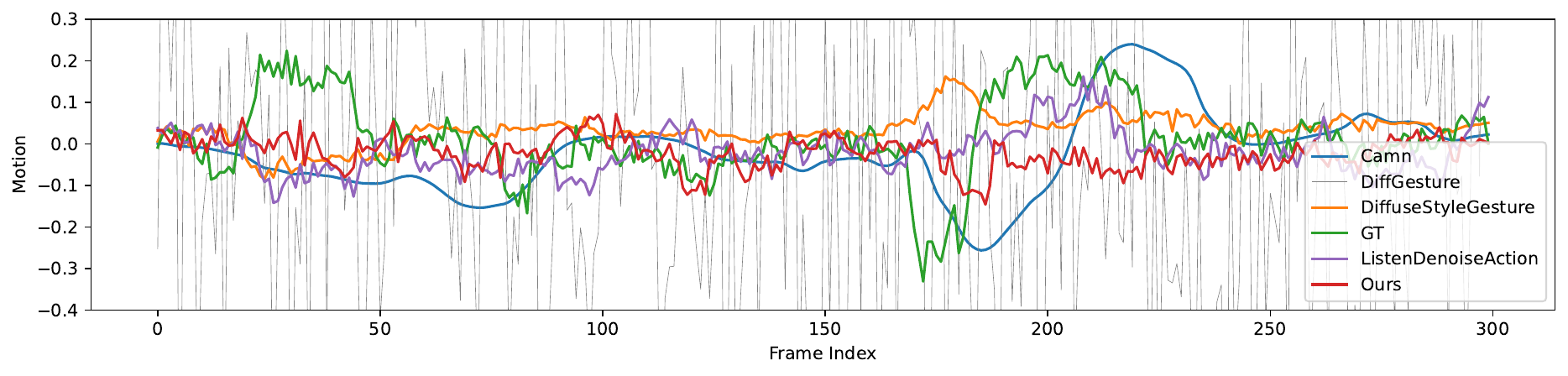}
   \caption{}
   \label{fig:motion_beat} 
\end{subfigure}

\begin{subfigure}[t]{0.99\textwidth}
   \includegraphics[width=0.99\linewidth]{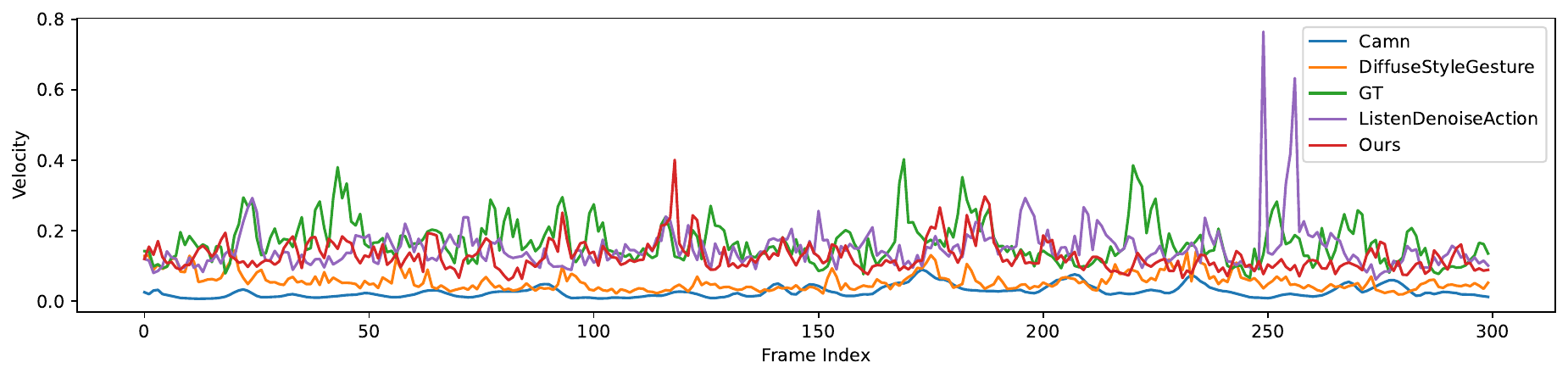}
   \caption{}
   \label{fig:velocity_beat}
\end{subfigure}

\begin{subfigure}[t]{0.99\textwidth}
   \includegraphics[width=0.99\linewidth]{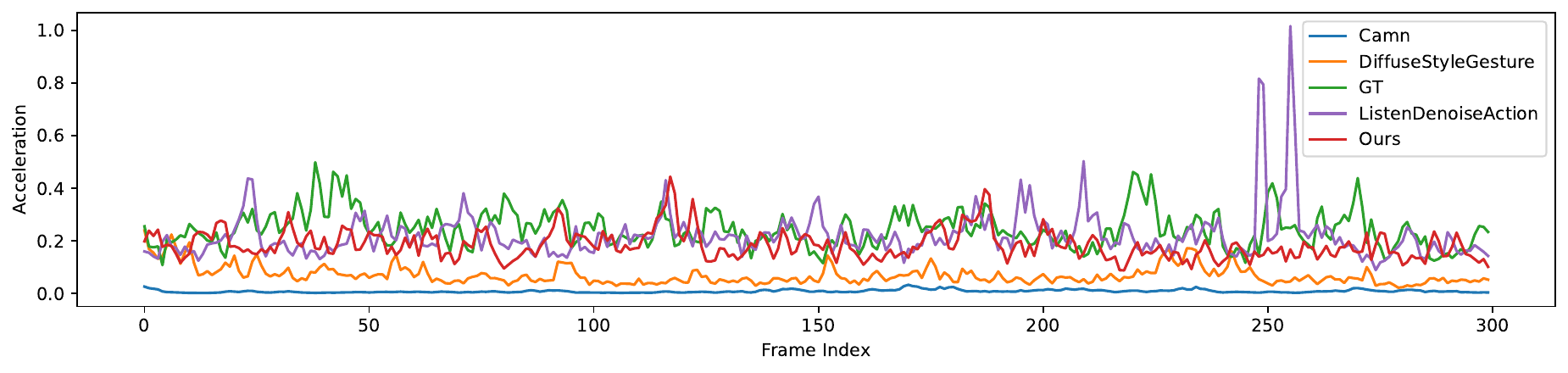}
   \caption{}
   \label{fig:acceleration_beat}
\end{subfigure}

\caption[Two numerical solutions]{\textbf{Motion, velocity, and acceleration comparison on BEAT.} The motion is from Scott3 in the BEAT test set. (a) The frame-wise channel mean of the normalized gesture parameters. (b) The velocity is computed as the frame-wise channel mean of the absolute residual of the adjacent frame motion. (c) The acceleration is computed as the frame-wise channel mean of the absolute residual of the normalized adjacent frame velocity. Due to the large jittering of DiffGesture, we only visualize it in the motion figure. }
\label{fig:agile_beat}
\end{figure*}

\noindent \textbf{Numerical value visualization.} In addition to watching the motion video to check the agility and smoothness, we also give a simple visualization of numerical values of motion, velocity, and acceleration. As shown in Figure~\ref{fig:velocity_beat} and Figure~\ref{fig:acceleration_beat}, we can obviously see that the velocity and acceleration of CaMN and DiffuseStyleGesture are consistently lower than GT, indicating lower agility. The DiffGesture has severe jittering gestures as shown in~\ref{fig:motion_beat}, therefore we did not visualize it in velocity and acceleration figures. Our method and ListenDenoiseAction have a comparative velocity as ground truth (GT) gestures, however, ListenDenoiseAction sometimes shows unstable motions such as at frame index of 195, 210, 250, and 260. As for smoothness, we can see CaMN has a very smooth curve in the motion Figure~\ref{fig:motion_beat}, however, it is too smooth compared to real motions. DiffuseStyleGesture and ListenDenoiseAction have comparative motion curve smoothness with ours and real motion, but if watch the motion video, we can still perceive that our method has the best smoothness compared to baselines, which is consistent with the results of user study.
\begin{table}[t]\footnotesize
  \begin{tabular}{l|c|c|c|c|c|c}
    \hline
               & GT    & Ours  & CaMN  & DSG   & LDA   & DG    \\ \hline
    AE $\downarrow$     & 0      & \underline{0.628} & \textbf{0.617} & 0.637 & 1.213 & 5.210  \\ \hline
    Vel $\dagger$  & 0.863 & \textbf{0.821} & 0.407 & 0.530 & 1.2531 & 3.399 \\ \hline
    MLVS$\dagger$ & 0.633 & \textbf{0.562} & 0.262 & 0.467 & 1.1493 & 3.992 \\ \hline
    \end{tabular}
    \caption{Smoothness and agility metric results. $\downarrow$: smaller is better. $\dagger$: closer to GT is better. }
    \label{tab:smoothness}
    \vspace{-0.5cm}
\end{table}

\section{Metrics Computation}

This section introduces the computational process of the metrics used in our experiment.
\begin{itemize}
    \item \textbf{Fr\'echet Motion Distance (FMD)} is slightly modified from Fr\'echet Gesture Distance~\cite{yoon2020speech}, which is a plausible metric consistent with human judgment. FMD calculates the Fr\'echet distance between the Gaussian mean and covariance of the latent feature distributions of synthesized motions and real motions. The latent features are extracted by a feature extraction network trained on the BEAT or SHOW dataset. Mathematically, 
   \begin{equation}
        FGD = |\mu_r - \mu_s|^2 + Tr(\sigma_r + \sigma_s - 2\sqrt{\sigma_r\sigma_s}),
    \end{equation}
    where $\mu_s$ and $\sigma_s$ are the mean and covariance of latent feature distribution of synthesized motions, and $\mu_r$ and $\sigma_r$ are the mean and covariance of the latent feature distribution of real motions. Fr\'echet Expression Distance (FED) and Fr\'echet Gesture Distance (FGD) are designed in a similar way. 

    \item \textbf{Percent of Correct Motions (PCM)} computes the percentage of correctly predicted motion results. In detail, we compute the generated motion parameter (i.e., body joint, expression weights) and its target is less than threshold $\delta$, then this parameter is considered to be correct. Specifically, PCM can be calculated as:
    \begin{equation}
        PCM = \frac{1}{T\times J}\sum^T_{t=1}\sum^J_{j=1}\mathbf{1}[|\hat{x}^j_t - x^j_t|_2 < \delta],
    \end{equation}
    where $\hat{x}^j_t$ and $x^j_t$ are the $j$-th joint of the generated motion and the ground-truth motion at the $t$-th frame, and $\mathbf{1}$ is the indicator function. We set $\delta=1.0$ on the SHOW dataset in our experiment.

    \item \textbf{Diversity metric (Div)} measures whether the generated motions are rich and diverse~\cite{bhattacharya2021speech2affectivegestures}. We denote the batch size at test time as $B$, then the \textbf{Div} is defined as:
    \begin{equation}
        Div = \frac{2}{B\times (B-1)}\sum^{B-1}_{i=1}\sum^{B}_{j=i+1}|\hat{x}_i-\hat{x}_j|_1,
    \end{equation}
    where $\hat{x}_i$ and $\hat{x}_j$ are the $i$-th and $j$-th generated motion sequence in a batch. We set $B=50$ in our experiment.

    \item \textbf{Beat Alignment (BA)} score is a metric for the alignment level between the audio and the generated motion. BA measures the average distance between each motion beat and its nearest corresponding audio beat. Mathematically, the BA score can be computed as
    \begin{equation}
        BA = \frac{1}{n}\sum^n_{i=1}\exp{-\frac{\min_{\forall b_j^a\in B^a} |b_i^m - b^a_j|^2}{2\sigma^2}},
    \end{equation}
    where $B^a = \{b_i^m\}$ denotes the motion beats, $B^a = \{b_i^m\}$ denotes audio beats. 
    For this metric, we follow the same implementation of BEAT~\cite{beat}~\footnote{https://github.com/PantoMatrix/BEAT} and TalkSHOW~\cite{talkshow}~\footnote{https://github.com/yhw-yhw/TalkSHOW} on their official GitHub repository.
\end{itemize}

\begin{figure*}[t]
\begin{center}
\includegraphics[width=1\linewidth]{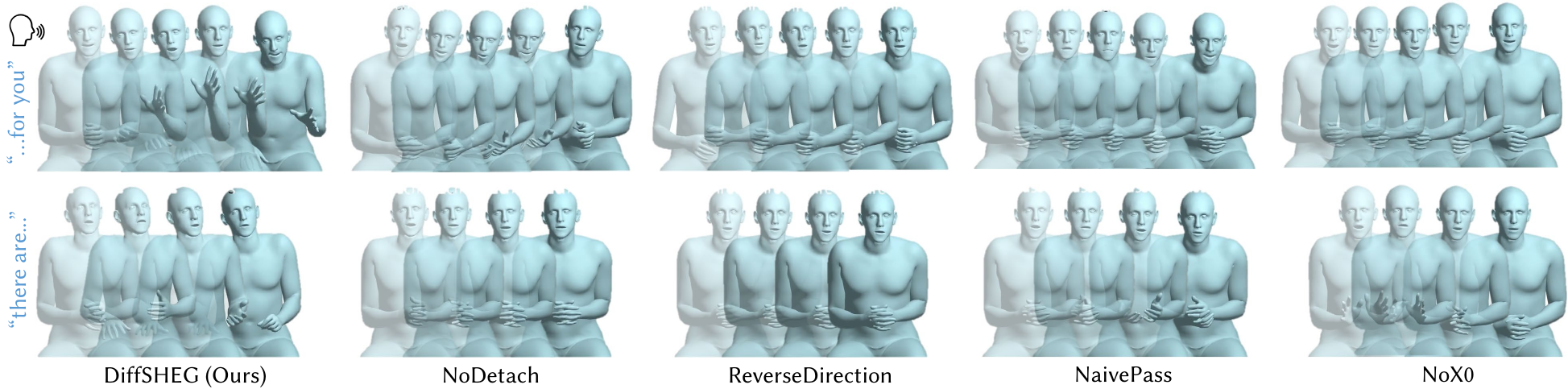}
\vspace{-0.7cm}
\end{center}
   \caption{Qualitative ablations. We compare our full method with ablations NoDetach, ReverseDirection, NaivePass, and NoX0 as described in Section~\ref{sec:ablation}. Our full version of DiffSHEG shows more diversified and meaningful gestures aligned with speech, while other ablations remain still or relatively small head and hand motions.}
\label{fig:qualitative_ablation}
\end{figure*}

\begin{table*}[t]
\centering
\begin{tabular}{@{}l|c|c|c|c|c|c|c@{}}
\hline
 & \multicolumn{1}{c|}{Motion} & \multicolumn{3}{c|}{Expression} & \multicolumn{3}{c}{Gesture} \\ \hline
             Methods    & FMD$\downarrow$          & FED$\downarrow$          & Lip-MSE$\downarrow$     & Lip-L1$\downarrow$             & FGD$\downarrow$       & BA$\dagger$        & SRGR$\uparrow$             \\ \hline
Ours             & \textbf{324.67}       & \underline{331.72}       & \textbf{0.1185}      & \textbf{0.9400}          & \textbf{438.93}    & \textbf{0.9139}    & \textbf{0.2507}        \\
w/o Hubert         & 347.07       & \textbf{325.91}       & 0.1612      & 1.1079         & 495.86    & 0.8916    & 0.2228        \\
w/o Mel            & 393.47       & 396.60       & 0.1297      & 0.9935         & 485.10    & 0.8933    & 0.2071        \\
w/o Mid            & 437.23       & 450.85       & 0.1193      & 0.9464         & 613.86    & 0.8933    & 0.2210        \\
UseFinal$\hat{x}_{0(0)}^E$ & 329.82 & 344.63 & 0.1627  & 1.1205      & 464.86    & 0.8760    & 0.2178       \\ \hline
\end{tabular}
\caption{Audio encoder ablation results. $\downarrow$: smaller is better. $\dagger$: closer to GT is better. 
}
\label{tab:rebuttal_ablation}
\vspace{-0.4cm}
\end{table*}

\begin{table}[t]
\centering
    \begin{tabular}{@{}lccccc@{}}
    \hline
          & Ours & CaMN & DG & DSG & LDA \\ \hline
    MSE$\downarrow$     & \textbf{0.1185}        & 0.2055        & 0.1783              & 0.2061                      & 0.1642                      \\ \hline
    L1$\downarrow$ & \textbf{0.9400}       & 1.2931        & 1.2002              & 1.3077                      & 1.1383                      \\ \hline
    \end{tabular}
    \vspace{-0.2cm}
    \caption{Lipsync results. $\downarrow$: smaller is better.}
    \label{tab:rebuttal_lipsync}
    \vspace{-0.3cm}
\end{table}

\section{Additional Implementation Details}
\textbf{Baseline Implementation.} We follow the official GitHub repositories of baselines for comparison. During the implementation, we found that LDA\cite{alexanderson2023listen} suffers from unstable sampling and may collapse to infinity numerical value during the gesture inference. In order to have a good Frechet distance score for LDA, we repair the collapsed motions by substituting them with the nearest motion clips. The metric values of LDA in Table~\ref{tab:table1} are computed on the repaired motion sequences.

\noindent\textbf{Fr\'echet Distance.} We re-train three Autoencoders for gesture-expression, gesture, and expression in order to compute their embeddings in neural feature space, where the gesture rotation is represented as axis-angle. We test Fr\'echet distances on un-overlapping clips to reflect the real distribution approximation ability of our Transformer-based diffusion model. The VAE architecture follows the implementation of BEAT~\cite{beat}.

\section{Additional Experiments}
\textbf{Qualitative ablation.} 
As shown in Figure~\ref{fig:qualitative_ablation}, compared to other baselines, our full version of DiffSHEG usually exhibits more diversified and meaningful gestures, including head motions. For example, when saying ``... for you", our full version of DiffSHEG generates the motion with a large stress gesture, while other ablations only show a small motion. When saying ``there are internal visions ...", our full version of motion shows a overturn and stop of hand as well as a turn of head, which corresponds to the speech stress on the word ``internal". In contrast, other ablations have hands remain still or relatively small motions.

\noindent\textbf{Audio encoder ablation.} 
As shown in Table~\ref{tab:rebuttal_ablation}, our final version gets overall better results with three ablations: removing the Hubert feature (w/o Hubert), removing Mel Spectrogram (w/o Mel), and removing the trainable audio Transformer which learns the mid-level features (w/o Mid).
The results indicate that removing any one of the three components will lead to a performance drop, especially the learnable mid-level audio transformer encoder. In addition, it seems that the low-level audio feature (Mel) has more influence on expression and the high-level audio feature (Hubert) affects more on gesture metrics.

\noindent\textbf{Lip synchronous metric.} 
We compute the \textbf{lip-sync error} in Table~\ref{tab:rebuttal_lipsync}. According to the results, our DiffSHEG achieves better lip-sync scores than all the baselines.

\section{Limitations}
Our method generates motions that faithfully imitate the training data distribution, therefore, high-quality real data is required to generate quality results. If the training data is noisy or has jittering or outlier motions, our generative model will also learn this pattern and generate similar noisy data. For example, we find that the female characters in BEAT have some jittering motion in training data, therefore our generated motion sometimes shows such jittering. Similarly, in the SHOW dataset, there are also some distortions and discontinuities in expressions, which lead to a similar pattern in the generated data.


\end{document}